\documentclass[twocolumn,showpacs,preprintnumbers,amsmath,amssymb, superscriptaddress, prl]{revtex4-1}

\usepackage{stmaryrd}
\usepackage{txfonts}
\usepackage{amssymb}
\usepackage{mathrsfs}
\usepackage{graphicx}% Include figure files
\usepackage{dcolumn}% Align table columns on decimal point
\usepackage{bm}% bold math
\usepackage{epsfig}
\usepackage{color}                    % For creating coloured text and background
\usepackage{hyperref}                 % For creating hyperlinks in cross references

\begin{document}

\preprint{\href{http://link.aps.org/doi/10.1103/PhysRevLett.110.087003}{S. Z. Lin and L. N. Bulaevskii, Phys. Rev. Let. {\bf 110}, 087003 (2013).}}

\title{Dissociation transition of a composite lattice of magnetic vortices in the flux-flow regime of two-band superconductors}
\author{Shi-Zeng Lin}\email{szl@lanl.gov}
\affiliation{Theoretical Division, Los Alamos National Laboratory, Los Alamos, New Mexico 87545, USA}

\author{Lev N. Bulaevskii}
\affiliation{Theoretical Division, Los Alamos National Laboratory, Los Alamos, New Mexico 87545, USA}

\begin{abstract}
In multiband superconductors, each superconducting condensate supports vortices with fractional quantum flux. In the ground state, vortices in different bands are spatially bounded together to form a composite vortex, carrying one quantum flux $\Phi_0$. Here we predict dissociation of the composite vortices lattice in the flux flow state due to the disparity of vortex viscosity and flux of the vortex in different bands. For a small driving current, the composite vortices starts to deform, but the constituting vortices in different bands move with the same velocity. For a large current, composite vortices dissociate and vortices in different bands move with different velocities. The dissociation transition shows up as increase of flux flow resistivity. In the dissociated phase, Shapiro steps are developped when an ac current is superimposed with a dc current.
\end{abstract}

%checked for Dissociation of composite vortex in two-band superconductors

\date{\today}

\maketitle
Multiband superconductivity is realized in many superconductors, such as the well studied $\rm{V_3Si}$, $\rm{Nb_2Se}$, the recently discovered $\rm{MgB_2}$ \cite{Nagamatsu01} and iron-based superconductors\cite{Kamihara08}. In these superconductors, electrons in different bands are cooled into distinct superconducting condensates, which interact with each other through interband tunneling. Multiband superconductivity may also exist in the proposed liquid hydrogen under pressure where both the proton and electron bands contribute to superconductivity\cite{Ashcroft68,Jaffe81}. In this case the interband tunneling is absent. The multiband superconductors attract considerable attention recently and phenomena unique to multiband superconductors, which do not have counterpart in single-band superconductors, have been predicted \cite{Leggett66,Tanaka02,Babaev02,Gurevich03, Gurevich03b, Babaev05,Smorgrav05, Gurevich06,Komendova2012,Lin2012PRL} and observed experimentally\cite{Bluhm06,Blumberg07,Moshchalkov09}. 

In multiband superconductors, there are several superconducting gaps $|\Delta_{\mu}|\exp(i \theta_{\mu})$, characterizing the quasiparticle excitation for superconducting condensate in each band respectively. Because the gap function is a complex function, each condensate thus supports vortex excitation with fractional quantized flux.\cite{Babaev02} The energy of a single fractional quantized vortex diverges logarithmically  due to the counter flows of different condensates that have no charge transfer and are not coupled with magnetic fields. Thus it is thermodynamically unstable. Stable fraction quantized vortex is predicted to exist in mesoscopic superconductors\cite{Chibotaru10,Pina12} or near the surface of multiband superconductors.\cite{Silaev11}

Vortices in different condensates appear simultaneously when an external field is applied to multiband superconductors. There are both interband and intraband vortex interaction. Vortices with the same polarization in the same condensate repel each other through the exchange of massive photon.  The vortices in different condensates interact repulsively due to the magnetic interaction. Meanwhile they attract each other due to the coupling to the same gauge field. The latter is more important and the net interaction of vortices in different condensates is attractive. They also attract with each other due to the interband tunneling in superconducting channel. Therefore vortex in different condensates in the ground state is bounded and their normal cores are locked together to form a composite vortex with the standard integer quantum flux $\Phi_0=hc/(2e)$. It is an interesting question whether the composite vortex can dissociate. The melting of composite vortex lattice due to thermal fluctuations was studied, and it was found that the vortex lattice in the superconducting condensate with weaker phase rigidity melts first as temperature increases, while the lattice ordering in other condensates remains.\cite{Smorgrav05}

\begin{figure}[t]
\psfig{figure=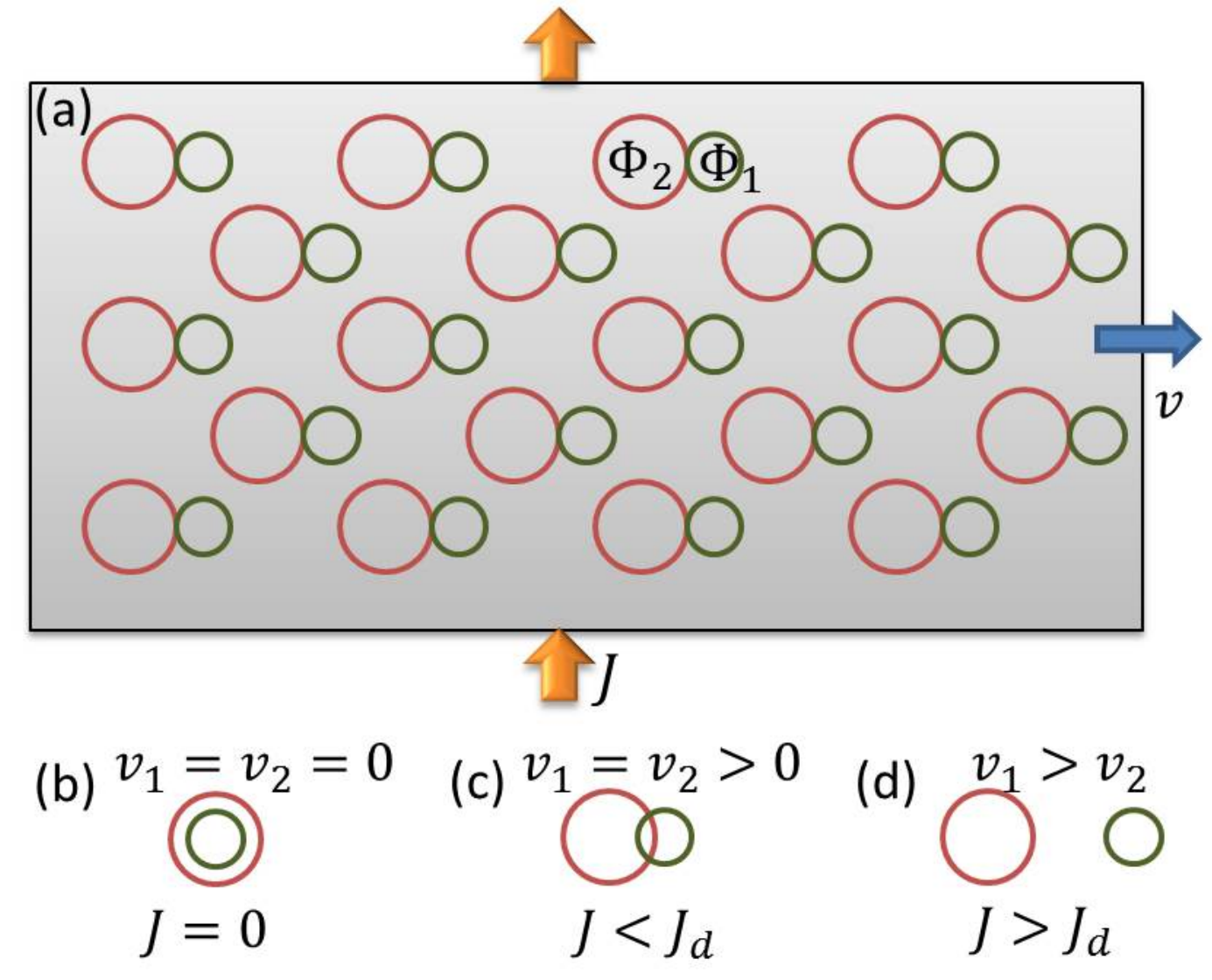,width=\columnwidth}
\caption{\label{f1}(color online) (a) Schematic view of vortex lattice in two-band superconductors. (b): In the ground state, the normal core of vortex in one condensate with flux $\Phi_2$ (red circle) is locked with that in the other condensate with flux $\Phi_1$ (green circle) to form a composite vortex with one quantum flux $\Phi_0=\Phi_2+\Phi_1$. (c): With a small current, two vortex lattices are shifted with respect to each other but move with the same velocity and the composite vortex is deformed. (d) At a large current, the two vortex lattices are decoupled, move with different velocities, and the composite vortex is dissociated. The dissociation transition occurs at $J_d$.}
\end{figure}

Here we consider the dissociation of composite vortex lattice in the flux flow region. With an external current, vortex in condensate with larger flux experiences stronger Lorentz force. On the other hand, the viscosity of vortex in different condensates is different due to the different size of normal core. As a result, vortices in some bands tend to move faster. For a small external current, the disparity of vortex motion can be balanced by the interband attraction between vortices in different condensates, and vortices in different condensates move with the same velocity, as shown in Fig. \ref{f1}. However at a large current, the interband attraction may be insufficient to compensate the disparity of driving force and viscosity of vortex in different condensates and composite vortices are dissociated, i.e. vortices in different condensates move with different velocities. In the decoupled phase, the flux flow resistivity increases. The Shapiro steps are induced under an additional ac current when the oscillation of vortex lattice induced by the ac current is resonant with the vortex oscillations due to the periodic potential created by the relative motion of vortex lattice in different condensates.

We use the London free energy functional for two-band superconductors with a Josephson-like interband coupling. The free energy density therefore can be written as
\begin{equation}\label{eq1}
\mathcal{F}_L=\frac{1}{8\pi }\sum _{\mu=1}^2\left[\frac{1}{\lambda _{\mu}^2}\left(\mathbf{A}-\frac{\Phi _0}{2\pi }\nabla \theta _{\mu}\right)^2+(\nabla \times \mathbf{A})^2\right]-\gamma\cos \left(\theta _1-\theta _2\right)
\end{equation}
where $\lambda_{\mu}=\sqrt{(m_{\mu} c^2)/(4\pi n_{\mu} e^2)}$ is the London penetration depth for each condensate with superfluid density $n_{\mu}$. $\mathbf{A}$ is vector potential, $m_{\mu}$ is the electron mass in $\mu$-th band and $\gamma$ is the interband Josephson coupling. The effective penetration depth for the two-band system is $\lambda^{-2}=\sum_{\mu=1}^2 \lambda_{\mu}^{-2}$. In the absence of external magnetic fields, the phase of different superconducting condensates is locked, $\theta_1=\theta_2$ for $\gamma>0$ and $\theta_1=\theta_2+\pi$ otherwise. The decoupling of phase due to the injection of current in superconducting wire both in equilibrium\cite{Gurevich06} and out of equilibrium \cite{Gurevich03} was discussed by Gurevich and Vinokur recently.

Minimizing $\mathcal{F}_L$ with respect to $\mathbf{A}$, we obtain the London equation
\begin{equation}\label{eq2}
\lambda ^2\nabla \times \nabla \times \mathbf{B}+\mathbf{B}=\Phi _{\mu}\sum _{\mu, j}\delta \left(\mathbf{r}-\mathbf{r}_{\mu,j}\right),
\end{equation}
where $\Phi _{\mu}=\lambda ^2\Phi _0/\lambda _{\mu}^2$ is the fractional quantum flux  and $\mathbf{r}_{\mu, j}=(x_{\mu, j}, y_{\mu, j})$ is the vortex coordinates for the vortex in the $\mu$-th condensate. The vortex line is assumed to be straight. $\mathcal{F}_L$ can be splitted into two contributions\cite{Silaev11} $\mathcal{F}_L=\mathcal{F}_m+\mathcal{F}_c$ with the magnetic coupling
\begin{equation}\label{eq3}
\mathcal{F}_m=\frac{1}{8\pi }\left[\mathbf{B}^2+\lambda ^2(\nabla \times \mathbf{B})^2\right].
\end{equation}
$\mathcal{F}_m$ is the same as that in single-band superconductors because there is only one gauge field $\mathbf{A}$ in superconductors. $\mathcal{F}_m$ accounts for the magnetic coupling between vortices in different condensates. $\mathcal{F}_c$ represents the coupling due to the phase difference between condensates
\begin{equation}\label{eq4}
\mathcal{F}_c=\frac{\Phi _1\Phi _2}{32\pi ^3\lambda ^2}\left[\nabla \left(\theta _1-\theta _2\right)\right]^2-\gamma \cos \left(\theta _1-\theta _2\right).
\end{equation}
$\mathcal{F}_c$ does not depend on $\mathbf{A}$ and accounts for the locking of two superconducting phases. For a fractional vortex where $\theta_1$ changes by $2\pi$ around $\mathbf{r}_{0}$ while $\theta_2$ does not change, the self-energy per unit length is
\begin{equation}\label{eq5}
E_{fv}=\left(\frac{\Phi _1}{4\pi  \lambda }\right)^2\ln \left(\frac{\lambda }{\xi _1}\right)+\frac{\Phi _1\Phi _2}{16\pi ^2\lambda ^2}\ln \left(\frac{L}{\xi _1}\right)+|\gamma|\int dr^2[1-\cos(\theta_1)]
\end{equation}
where $L$ is the linear size of the system and $\xi_{\mu}$ is the coherence length. $E_{fv}$ diverges at $L\rightarrow \infty$ due to the neutral mode described by the term proportional to $[\nabla(\theta_1-\theta_2)]^2$ in Eq. (\ref{eq4}). The Josephson contribution in Eq. (\ref{eq5}) renders the fractional vortex linearly divergent in $L$. Thus a fractional vortex is thermodynamically unstable in bulk superconductors \cite{Babaev02}. For a composite vortex where $\theta_{\mu}$ changes by $2\pi$ around the same position, $\mathcal{F}_c=0$ and its self-energy is finite.

To calculate intraband and interband interaction for vortex, one needs to know $\theta_{\mu}$. They can be obtained by minimizing Eq. (\ref{eq4}) with respect to $\theta_{1}$,
\begin{equation}\label{eq6}
\frac{\Phi _1\Phi _2}{16\pi ^3\lambda ^2}\nabla ^2\left(\theta _1-\theta _2\right)-\gamma  \sin \left(\theta _1-\theta _2\right)=0,
\end{equation}
subject to the boundary condition accounting for vortices
\begin{equation}\label{eq6a}
\nabla\times(\nabla \theta_{\mu})=2\pi\sum_{\mu, j}\delta(\mathbf{r}-\mathbf{r}_{\mu, j}).
\end{equation}
The interaction between vortices according to Eqs. (\ref{eq6}) and (\ref{eq6a}) is nonlinear, thus it is many-body interaction. The term $\nabla^2(\theta_1-\theta_2)$ is of the order $1/\bar{a}^2$ with $\bar{a}$ being  the average distance between vortices in the same condensate.  For a strong field such that  $\bar{a}\ll\lambda_J=\sqrt{{\Phi _1\Phi _2}/({16\pi ^3\lambda ^2|\gamma|}})$, the sine term becomes $\bar{a}^2/\lambda_J^2$ times smaller than the gradient term thus can be neglected. In this case, $\mathcal{F}_c$ reduces to the free energy for the $XY$ model. For $\rm{MgB_2}$, $\gamma\approx 150\ \rm{J/m^3}$\cite{Gurevich03b}, it requires fields stronger than $4$ T at temperature $T=0$ K. For $\rm{V_3Si}$\cite{Kogan09} and $\rm{FeSe}_{1-x}$\cite{Khasanov10}, the required field is smaller because the interband coupling $\gamma$ is much weaker.

We then discuss the interband and intraband interaction between vortices, neglecting the Josephson interband coupling term. Then both $\mathcal{F}_m$ and $\mathcal{F}_c$ are quadratic in $\mathbf{B}$ and $\theta_{\mu}$, therefore the interaction between vortices is pairwise. $\mathcal{F}_m$ accounts for short-range interband and intraband repulsion between vortices with the same polarization. $\mathcal{F}_c$ describes the plasma interaction in two dimensions and the interaction between vortices is long range. The term proportional to $(\nabla\theta_{\mu})^2$ accounts for intraband repulsion and the term proportional to $-\nabla\theta_1\nabla\theta_2$ produces interband attraction between vortices in different condensates. This interband attraction outweighs the interband repulsion in $\mathcal{F}_m$. The intraband repulsion between two vortices in the same condensate separated by a distance $\mathbf{r}_{\mu, ij}\equiv\mathbf{r}_{\mu,i}-\mathbf{r}_{\mu,j}$ is
\begin{equation}\label{eq7}
V_{\rm{intra}}(r_{\mu, ij})=\frac{\Phi _{\mu}^2}{8\pi ^2\lambda ^2}K_0\left(\frac{r_{\mu, ij}}{\lambda }\right)-\frac{\Phi _1\Phi _2}{8\pi ^2\lambda ^2}\ln \left(r_{\mu, ij} \right),
\end{equation}
and the interband attraction between two vortices in the different condensates with a separation $\mathbf{r}_{12, ij}\equiv\mathbf{r}_{1,i}-\mathbf{r}_{2,j}$ is
\begin{equation}\label{eq8}
V_{\rm{inter}}({r}_{12, ij})=\frac{\Phi _1\Phi _2}{8\pi ^2\lambda ^2}\left[K_0\left(\frac{{r}_{12, ij}}{\lambda }\right)+\ln \left({r}_{12, ij}\right)\right].
\end{equation}
Equations (\ref{eq7}) and (\ref{eq8}) are valid away from vortex cores. 

In the flux flow state, vortices in each condensate driven by the Lorentz force move and cause dissipation, resulting in vortex viscosity. The dissipation is due to the motion of normal core, thus the viscosity for vortices in each band depends on $\xi_{\mu}$ of the corresponding band. In the framework of the Bardeen-Stephen model, the viscosity is given by $\eta_{\mu}=\Phi_0^2/(2\pi c^2\xi_{\mu}^2)$.  We use the quasistatic approximation, i.e. the vortex structure in each condensate does not change in the dynamic region, and introduce overdamped dynamics for vortices
\begin{align}\label{eq9}
\nonumber \eta_\mu\partial_t r_{\mu, i}=\frac{1}{8\pi ^2\lambda ^3}\sum _j\left[\Phi _{\mu }^2 K_1\left(\frac{r_{\mu ,ij}}{\lambda }\right)+\frac{\Phi _{1 }\Phi _{2 }\lambda}{r_{\mu, ij}}\right]\\
+\frac{\Phi _{1 }\Phi _{2 }}{8\pi ^2\lambda ^3}\sum _j\left[K_1\left(\frac{r_{12 ,ij}}{\lambda }\right)-\frac{\lambda }{r_{12, ij}}\right]+ \frac{J \Phi_{\mu}}{c}.
\end{align}
where $J$ is the external current. The effect of disorder becomes less important in the flux flow region because the disorder is averaged out by vortex motion, and lattice ordering is improved \cite{Koshelev94,Besseling03}. In the lattice phase, the intraband vortex interaction vanishes due to symmetry. The interband vortex attraction can be written in the momentum space and we only take the contribution from the dominant wavevector $\mathbf{G}=(\pm 2\pi/a, 0)$ for the vortex lattice moving along the $x$ direction. Here $a$ is the lattice constant and we assume a square lattice. In the region $2\pi \lambda/a\gg 1$, the equation of motion for the center of mass of vortex lattice $R_{\mu}$ in each band becomes
\begin{align}\label{eq10}
{\eta _2'}\partial _t\left(R_2-R_1\right)=-\left(1+{\eta _2'}\right)\sin \left(R_2-R_1\right)+\left({\Phi _2'}-{\eta _2'}\right)J,
\end{align}
\begin{equation}\label{eq11}
\partial _tR_1+{\eta _2'}\partial _tR_2=\left(1+{\Phi _2'}\right)J.
\end{equation}
We have written  Eq. (\ref{eq10}) in term of the relative motion between the vortex lattice in different bands, and the sine term accounts for the attraction between the two lattices. Here dimensionless units are introduced: length is in unit of $a/(2\pi)$, time is in unit of $\eta_1 a/ (2\pi F_d)$, current is in unit of $c F_d/\Phi_1$. $F_d$ is the maximum attractive force between two lattices $F_d={\Phi _1\Phi _2a}/({64\pi ^6\lambda ^4})$. $\Phi_2'\equiv\Phi_2/\Phi_1$ and $\eta_2'\equiv\eta_2/\eta_1$. Similar equations for vortex motion in bilayer superconducting films was presented in Ref. \cite{Clem74}. 

\begin{figure}[t]
\psfig{figure=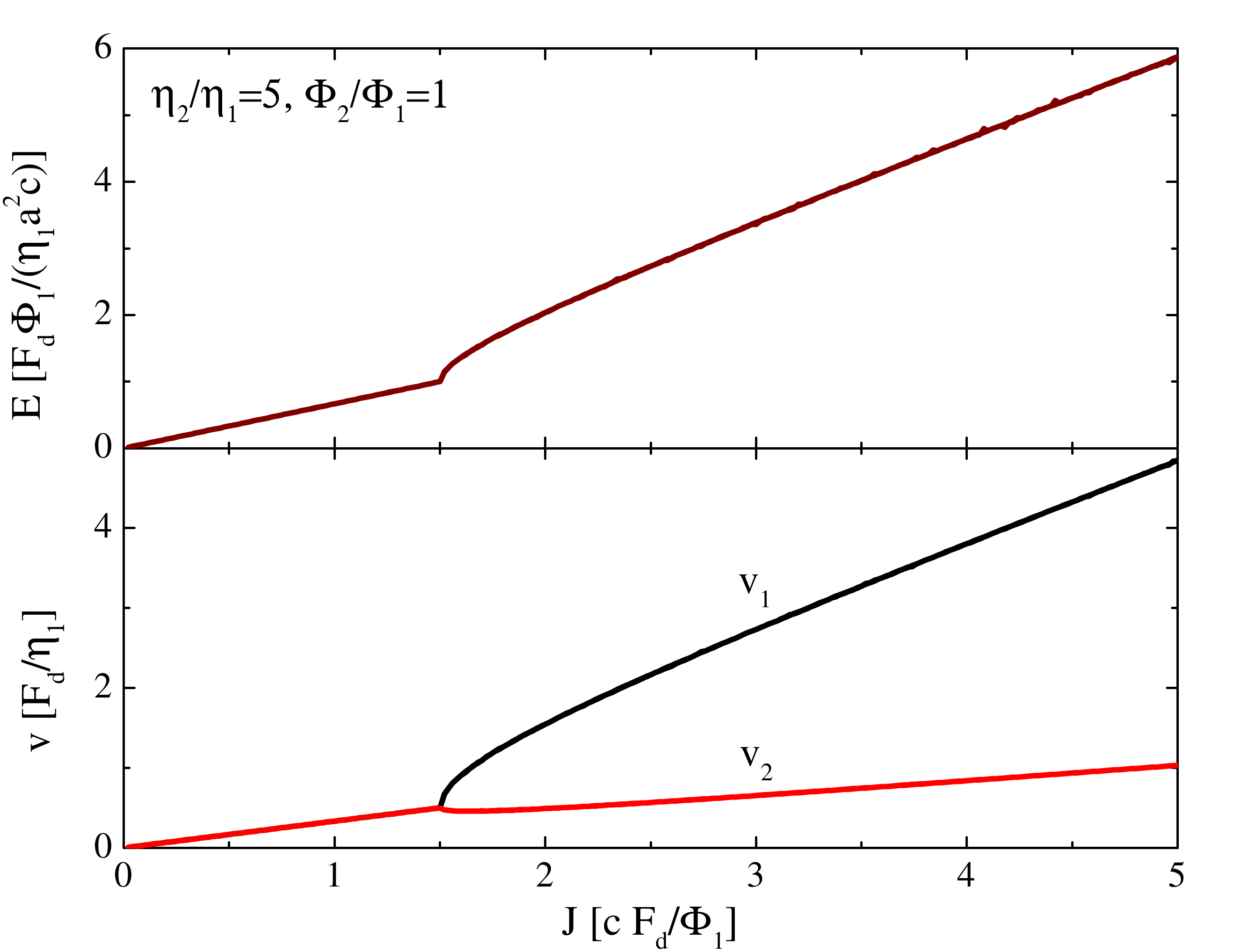,width=\columnwidth}
\caption{\label{f2}(color online) The dependence of velocity $v_1$, $v_2$ and electric field $E$ on the current $J$, obtained from Eqs. (\ref{eq10}), ~(\ref{eq11}) and (\ref{eq13}).}
\end{figure}

At a small current, two vortex lattices in different band move as a whole with a velocity $v_1=v_2=\left(1+{\eta _2'}\right)^{-1}\left(1+{\Phi _2'}\right)J$. The centers of mass of these two lattices deviate with a separation $\sin^{-1}\left[\left(1+{\eta _2'}\right)^{-1}\left({\Phi _2'}-{\eta _2'}\right)J\right]$. At this stage, the composite vortex starts to deform. The maximum attraction is reached at $R_2-R_1=\pi/2$ or $a/4$ in real unit. At a threshold current
\begin{equation}\label{eq12}
J_d=\left|\left(1+{\eta _2'}\right)\left({\Phi _2'}-{\eta _2'}\right)^{-1}\right|,
\end{equation}
these two lattices decouple and move at different velocities, resulting in the dissociation of composite vortex. Their corresponding velocity is
\begin{align}\label{eq13}
v_{\mu}=\left(1+{\eta _2'}\right)^{-1}\left[\left(1+{\Phi _2'}\right)J-\frac{\eta _1}{\eta _{\mu}}\sqrt{\left({\Phi _2'}-{\eta _2'}\right)^2J^2-\left(1+{\eta _2'}\right)^2}\right]
\end{align}
The dependence of $v_{\mu}$ on $J$ is displayed in Fig. \ref{f2}. At a large current $J\gg J_d$, each lattice behaves independent with velocities $v_1=J$ and $v_2=J /\eta_2'$.

The decoupling of two lattices can be observed experimentally in transport measurements. The I-V characteristics can be derived from power balance condition $\eta_1 v_1^2+\eta_2 v_2^2= J E a^2$ with $E$ the electric field. The I-V curve is shown Fig. \ref{f2}, where the differential resistivity $dE/dJ$ increases in the decoupled phase. Experimental observation of such increase may be challenging because the decoupling current usually is large, and the Larkin-Ovchinnikov instability of vortex lattice may be important\cite{Larkin76}. The vortex core shrinks due to the electric field caused by vortex motion and the flux flow resistivity increases with current, which blurs the dissociation transition in the I-V curve. Nevertheless, the dissociation transition can be confirmed unambiguously by measurement of the Shapiro steps in the decoupled phase, as discussed below.

 \begin{figure}[t]
\psfig{figure=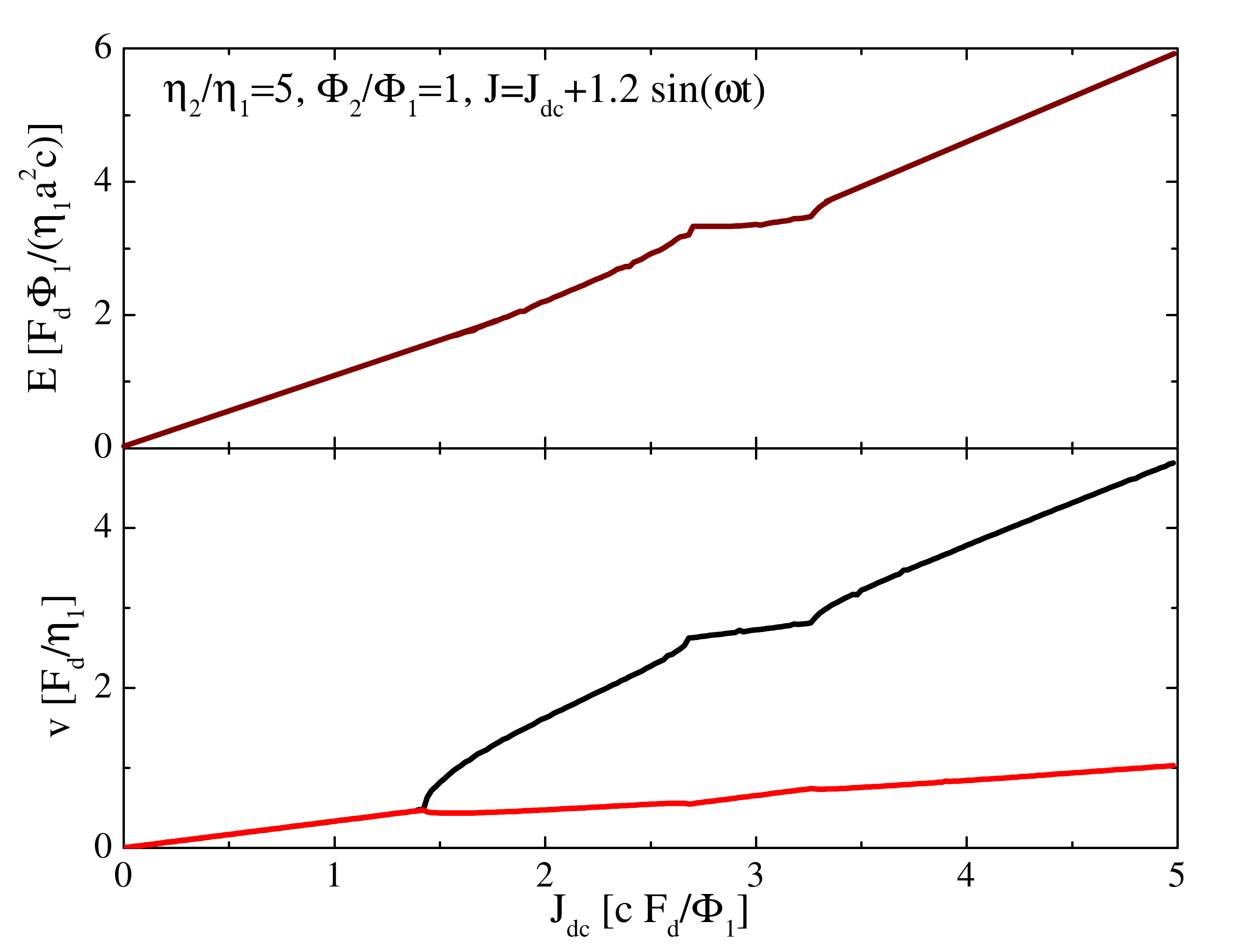,width=\columnwidth}
\caption{\label{f3}(color online) The same as Fig. \ref{f2}, but with an ac current in addition to a dc current.}
\end{figure}

In the decoupled phase, if one takes one lattice as reference, the other lattice experiences periodic potential induced by the reference lattice. When an ac current is added in superposition to the dc current, the oscillation of the moving lattice induced by the ac current may be in resonance with the oscillation due to the periodic potential of the reference lattice if the period of the ac current matches with the period of the potential. This results in Shapiro steps in I-V curves. The physics is the same as that in Josephson junctions and Eq. (\ref{eq10}) also describes the phase dynamics in overdamped Josephson junctions. With a current $J=J_{\rm{dc}}+{\rm{Re}} \left[J_{\rm{ac}}\exp[i (\omega t+\varphi)]\right]$, the center of mass of each lattice is $R_{\mu}={v}_{\mu}t+{\rm{Re}} \left[A_{\mu}\exp[i({v}_2-{v}_1)t]\right]$  in the region $|v_1-v_2|=\omega\gg 1$. From Eqs. (\ref{eq10}) and (\ref{eq11}), we obtain $A_{\mu}={\eta _{\mu}}[1-i{\Phi _{\mu}}J_{\text{ac}}\exp(i\varphi)/{\Phi _1}]/[{\left(v_2-v_1\right){\eta _1}}]$. The dc current is $J_{\text{dc}}=\left({\Phi _2'}-{\eta _2'}\right)^{-1}{\rm{Re}}\left[{\eta _2'}\left(v_2-v_1\right)+\left(1+{\eta _2'}\right){\left(A_2-A_1\right)}/{2}\right]$. When one changes $J_{\rm{dc}}$, $\varphi$ adjusts correspondingly because $v_1-v_2$ is locked with the driving frequency $\omega$, and a Shapiro step is traced out. The height of the Shapiro step is 
\begin{align}\label{eq14}
J_{sp}=\frac{\left(1+{\eta _2'}\right)\left({\eta _2'}{\Phi _2'}-1\right)}{({v_2-v_1})\left({\Phi _2'}-{\eta _2'}\right)}.
\end{align}
The results for Shapiro step obtained by solving Eqs. (\ref{eq10}) and (\ref{eq11}) with $J=J_{\rm{dc}}+1.2\sin(\omega t)$ numerically, are shown in Fig. \ref{f3}. Shapiro steps appears when $\omega=v_1-v_2$. Here we only considered the dominant resonance. There are also Shapiro steps at $n \omega=(v_1-v_2)$ with an integer $n>1$ with a smaller height. In the presence of quenched disorder, the Shapiro steps occurs at $n\omega=v_{\mu}$ (with the reduced units) due to the periodic passing of vortex lattice through the defects.\cite{Fiory71,Schmid73} These steps can be distinguished from those induced by relative motion of two vortex lattices, because their resonance condition is different.

The decoupling of composite vortex lattice depends on the two parameters $\Phi_2/\Phi_1$ and $\eta_2/\eta_1$. Generally the effect is present in all multiband superconductors. However for a small disparity between bands in $\xi_{\mu}$ and $\lambda_{\mu}$, the current  $J_d$ is high. There are superconductors with large disparity among condensates, such as the proposed liquid hydrogen superconductors due to large mass difference between proton and electron. To estimate $J_d$ for $\rm{MgB_2}$, we take $\xi_1=13$ nm, $\xi_2=51$ nm, $\lambda_1=47.8$ nm and $\lambda_2=33.6$ nm at $T=0$ K \cite{Moshchalkov09},  $\rho_{\mu}=10^{-9}\ \rm{\Omega\cdot m}$\cite{Xi08} and $a=40$ nm corresponding to field at $B\approx 1$ T. Then we obtain $J_d=5\times 10^9\ \rm{A/m^2}$, which is much smaller than the depairing current. The velocity of vortex lattice at dissociation transition is $v_1=v_2\approx 3\ \rm{m/s}$, which is smaller than the typical Larkin-Ovchinnikov instability velocity for vortex lattice.\cite{Doettinger94}  In the presence of defects, to observe the dissociation transition, one first needs to overcome the pinning potential, thus the effective dissociation current is the sum of the depinning current and the dissociation current $J_d$ for the clean system.

To illustrate the idea, we have employed the phenomenological London approach in Eq. (\ref{eq1}), which remains valid far away from the transition temperature $T_c$. For temperatures close to $T_c$, it was shown in Ref. \onlinecite{Kogan11} by using the standard Ginzburg-Landau (GL) model that there is only one coherence length in two-band superconductors and the systems behave as single-band superconductors. Thus the effect of dissociation of composite vortex lattice is absent near $T_c$. Far away from $T_c$, the standard GL model becomes inapplicable. Recently an extended GL model which is capable of describing different spatial lengths of condensate in different bands was derived from a microscopic model.\cite{Shanenko11,Vagov12,Komendova11} The extended GL model reduces to the London equation in Eq.(\ref{eq1}) if high order corrections are neglected and the amplitudes of superconducting order parameters are uniform in space.  The extended GL model is also a proper framework to discuss the vortex dissociation in multiband superconductors, from which the two phenomenological parameters $\lambda_{\mu}$ in Eq. (\ref{eq1}) may be calculated.

The dissociation of composite vortex lattice shares some similarity to that in multilayer superconductors\cite{Giaever65} and also in cuprate superconductors\cite{Busch92,Wan93,Safar94}. In the latter case, vortices in different layers both carry $\Phi_0$ flux and the dissociation occurs in the real space, while for multiband superconductors, vortices carry fractional flux and the dissociation occurs in the band (momentum) space. The decoupling transition has been observed experimentally\cite{Giaever65,Busch92,Wan93,Safar94} and discussed theoretically\cite{Cladis68,Clem74,Uprety95} decades ago. The Shapiro steps are also observed in the decoupled phase in multilayer superconductors\cite{Gilabert94}. These observations in multilayer superconductors corroborate the possible observation of the predicted dissociation of composite vortex in multiband superconductors. 

To summarize, we have predicted the dissociation  of composite vortices in two-band superconductors in the flux flow region because of the disparity in vortex viscosity and flux of vortex in different condensates. At a small velocity, the two vortex lattices are shifted with respect to each other and the composite vortices are deformed. At a high velocity, the two vortex lattices move with different velocities resulting in the dissociation of composite vortices. In the decoupled phase, the flux flow resistivity increases. The Shapiro steps are induced when an ac current is applied in addition to the dc current under an appropriate condition.

\vspace{2mm}

 \noindent {\it Acknowledgement --}
The authors are grateful to C. Reichhardt, E. Babaev, A. E. Koshelev and M. V. Milo\v{s}evi\'{c} for helpful discussions. This work was supported by the US Department of Energy, Office of Basic Energy Sciences, Division of Materials Sciences and Engineering.

%\bibliography{reference}

\end{document}